%% file: ec-mm-paper.tex
\documentclass[11pt, conference, onecolumn,  compsocconf ]{IEEEtran}

\usepackage[cmex10]{amsmath}
\usepackage{amsthm, amssymb}
\usepackage{cite}
\usepackage{array}
\usepackage{graphicx}
\usepackage[dvipsnames,usenames]{color}
\usepackage{xspace}

\def\doctype{1}

\def\tsubmission{2}
\ifnum\doctype=\tsubmission
	\newcommand{\full}[1]{}
	\newcommand{\submit}[1]{#1}
\else
	\newcommand{\full}[1]{#1}
	\newcommand{\submit}[1]{}
\fi

\newtheorem{theorem}{Theorem}
\newtheorem{lemma}[theorem]{Lemma}

\theoremstyle{definition}

\newcommand{\adwords}{{\sc AdWords}\xspace}
\newcommand{\gen}{{\sc AdGeneral}\xspace}
\newcommand{\lam}{{\sc AdLaminar}\xspace}
\newcommand{\genaon}{{\sc AdGen-AON}\xspace}
\newcommand{\genp}{{\sc AdGen-P}\xspace}

\newcommand{\opt}{{\text{\sc opt}}\xspace}
\newcommand{\algo}{{\text{\sc algo}}\xspace}

        {\hspace*{\fill}$\Box$\par}
        
\newcommand{\eat}[1]{}

\begin{document}



\title{Online Budgeted Allocation with General Budgets}
\author{\IEEEauthorblockN{Nathaniel Kell \hspace*{20pt} Debmalya Panigrahi}
\vspace*{2pt}
\IEEEauthorblockA{Department of Computer Science,
Duke University,
Durham, NC, USA.\\
{\tt Email: \{kell,debmalya\}@cs.duke.edu}}}

\maketitle

\begin{abstract}
We study the online budgeted allocation (also called \adwords) problem, where a set of impressions arriving online are allocated to a set of budget-constrained advertisers to maximize revenue. Motivated by connections to Internet advertising, several variants of this problem have been studied since the seminal work of  Mehta, Saberi, Vazirani, and Vazirani (FOCS 2005).  However, this entire body of work focuses on a single budget for every advertising campaign, whereas 
in order to fully represent the actual agenda of an advertiser, an advertising budget should be expressible over multiple tiers of user-attribute granularity.  
A simple example is an advertising campaign that is constrained by an overall budget but is also accompanied by a set of sub-budgets for each target demographic.
In such a contract scheme, an advertiser can specify their true user-targeting goals, allowing the publisher to fulfill them through relevant allocations.



In this paper, we give a complete characterization of the \adwords problem for general advertising budgets. In the most general setting, we show that, unlike in the single-budget \adwords problem, obtaining a constant competitive ratio is impossible and give asymptotically tight upper and lower bounds.  However for our main result, we observe that in many real-world scenarios (as in the above example), multi-tier budgets have a {\em laminar} structure, since most relevant consumer or product classifications are hierarchical. For laminar budgets, we obtain a competitive ratio of $e/(e-1)$ in the small bids case, which matches the best known \adwords result for single budgets. Our algorithm has a primal-dual structure and generalizes the primal-dual analysis for single-budget \adwords first given by Buchbinder, Jain, and Naor (ESA 2007). However many new ideas are required to overcome the barriers introduced by laminar budgets---our algorithm uses a novel formulation that overcomes non-monotonicity in the syntactically defined dual variables, as well as a dynamically maintained labeling scheme that properly tracks the ``most-limiting" budgets in the hierarchy. 



\end{abstract}


\thispagestyle{empty} \setcounter{page}{0} 

\input ec-mm-introduction.tex


\section{\adwords with laminar budget constraints (\lam)}
\label{sec:lam-sb}
Recall the \lam problem: we given a set of offline bidders $U$ and a set of impressions $V$ that arrive online, 
where each bidder-impression pair $(u,v)$ is specified by a bid value $r_{uv}^{(k)}$ for each dimension $k \in K$.
The revenue generated from a bidder $u$ is subject to an arbitrary set of budget
constraints $S_u$, where each constraint $s \in S_u$ caps the total revenue generated from a 
subset of dimensions $K_s$ to a given budget $B_u^{(s)}$.  We assume that the sets 
$\{K_s : s \in S_u\}$ form a laminar family, i.e., for every pair 
of intersecting sets in $\{K_s : s \in S_u\}$, one is contained in the other.
  
In this section, we give an algorithm for the \lam problem with a competitive ratio of $e/(e-1)$ 
under the small bids assumption (Theorem~\ref{thm:lam}).  This bound is tight because of a matching 
lower bound for the \adwords problem~\cite{MehtaSVV07}. Throughout this section, 
we assume that for every dimension $k \in K$, a constraint 
$s$ with $K_s = \{k\}$ appears in $S_u$ for each bidder $u$. 
This is wlog since a budget can be made arbitrarily large. 
We will call these constraints {\em singleton budgets} of bidder $u$.  

\input ec-mm-laminar-small.tex

\section{\adwords with general budget constraints (\gen)}
\label{sec:gen}

\input ec-mm-general.tex

\bibliographystyle{plain}
\bibliography{ref}

\clearpage

\appendices
\setcounter{section}{0}

\input ec-mm-gen-moderate-lb.tex

\input ec-mm-gen-small-lb.tex

\input ec-mm-gen-moderate-algo.tex

\end{document}

%% file: ec-mm-introduction.tex
\section{Introduction}
\label{introduction}

The online budgeted allocation problem, also called the \adwords problem, 
has had a significant impact on the theory and practice of online (Internet) advertising. 
In this problem, an advertisement publisher is tasked with matching user-generated advertisement 
slots on a web page (typically called {\em impressions}) to advertisers (typically called {\em bidders}).
More formally, the publisher is given a set of offline bidders $u \in U$ at the outset of the problem, 
each of which is specified by a budget $B_u$ 
indicating the maximum revenue the publisher can receive from bidder $u$. A set of impressions $v \in V$ 
then arrive in an online sequence and must each be irrevocably assigned to a unique bidder. 
Upon assigning impression $v$ to bidder $u$, the publisher receives revenue $r_{uv}$ (typically 
called a {\em bid value}) from bidder $u$. The objective is to maximize the total revenue 
generated over all impressions. 

The \adwords problem was introduced in the seminal work of Mehta, Saberi, Vazirani, and
Vazirani~\cite{MehtaSVV07}. While this problem generalizes the classic {\em online matching} problem
introduced by Karp, Vazirani, and Vazirani~\cite{KarpVV90}, the research focus has been on the 
so called {\em small bids} case, i.e., on algorithmic performance as the 
ratio $\frac{\max_{v\in V} b_{uv}}{B_u}$ tends to 0 for all advertisers. This assumption models 
most real-world scenarios, where the revenue generated from 
a single impression is infinitesimal compared to the total budget of an advertiser. 
The small bids assumption distinguishes \adwords from online matching and makes them incomparable
from a technical perspective; given its natural applicability and popularity in Internet 
advertising, we will also focus primarily on the small bids case.

Since the introduction of these problems, several variants of \adwords and online matching have been studied, motivated primarily
by the evolving challenges advertisement publishers face in practice. For example, 
Agrawal and Devanur~\cite{AgrawalD15} recently considered arbitrary linear 
and non-linear convex budget constraints for the stochastic input 
model,  and Devanur and Jain \cite{DevanurJ12} studied concave returns on revenue, 
both motivated by problem features such as under-delivery penalties and pay-per-click advertisements.
Another motivation for these variants that is receiving increasing attention
is that of {\em impression diversification} or {\em representativeness}. More specifically, a campaign contract 
is usually of the form: ``deliver ten million advertisements to Californian females in the month of July." Although such an agreement 
clearly indicates a target group to which the publisher should restrict its assignments, often the
advertiser still wants the impressions to be equally spread among the sub-populations of the targeted group (e.g., 
in the above contract, the advertiser will likely be unhappy if their ads are only shown to white females in their twenties living in Los Angeles).  
Although it can be at odds with short-term revenue gains, ensuring diversity is crucial with respect to {\em long-term} revenue for the publisher, as
advertisers that see a high return on investment (in this case via reaching desired audiences) are more likely to continually purchase future contracts. 
 
There are several recent works that address impression diversification (e.g., see \cite{GhoshMPV09}, \cite{BharadwajCMNTVVY12}, and \cite{HojjatTCY14}).
In many of these results, the objective function of the problem incorporates a diversity penalty, usually 
in the form of a distance function that incurs a cost if the algorithm's assignment differs too much from the advertiser's ideal allocation. 
However another natural approach, which to the best of our knowledge has yet to be considered, is a scheme where contracts are specified 
over multiple tiers of user-attribute granularity.  Recalling our previous contract example, in a multi-tier scheme the advertiser 
could further specify constraints in terms of age groups, e.g., 
``no more than four million of the total ten million advertisements should be shown to age groups 20-29, 30-39, and 40+, respectively." Further constraints 
could also be placed on each age group in terms of residency, e.g., 
``of the maximum four million advertisements assigned in each age group, no more than two million should be shown 
to residents in Los Angeles."  This scheme has several appealing features, the foremost being that it allows advertisers to explicitly indicate their true user-targeting goals 
with a high degree of expressibility (whereas penalty functions often assume that the ideal allocation must follow the same distribution as the overall targeted population, i.e., 
is forced to be a representative sample). 


\begin{figure} 
\begin{center}
\includegraphics[scale=0.7]{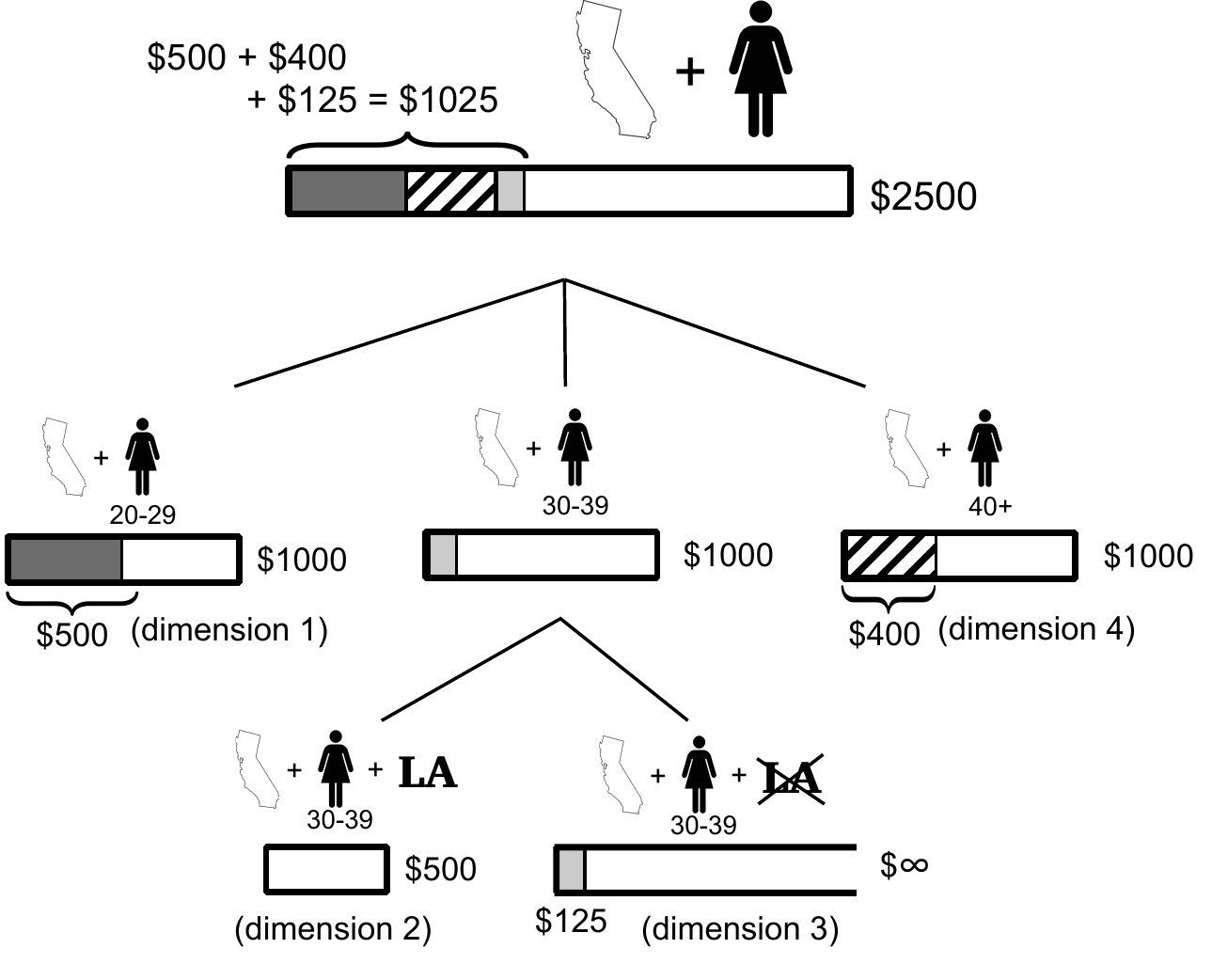}
\caption{{\small Illustration of how the ``Californian female" example translates into a bidder's budget in an instance of \gen (and in 
this case also \lam).  For simplicity, we only have a Los Angeles residency subdivision for the 30-39 age range, giving us 
four dimensions in total. Observe that $K_s = \{2,3\}$ for the 30-39 age-range budget, and 
and $K_s = \{1,2,3,4\}$ for the overall budget; furthermore, we have that $B_u^{\{2,3\}} =  1000$ and $B_u^{\{1,2,3,4\}} = 2500$. 
The algorithm has currently earned \$500, \$100, and \$400 on dimensions 1, 3, and 4, respectively (and \$0 on dimensions 2); therefore, 
we have used \$1025 out of the overall budget's capacity of \$2500. If an impression $v$ is newly assigned to this bidder $u$ such that 
$r_{uv}^{(1)} = \$10$ and $r_{uv}^{(k)} = 0$ for $k =2,3,4$, then we will have then earned \$510 on dimension 1 and \$1035 overall.}} 
\label{fig:probdef} 
\end{center}
\end{figure}

With this motivation, we introduce a generalization of the \adwords 
problem that we call {\em \adwords with general budgets} (denoted \gen). 
As earlier, there is an offline set of advertisers $U$, and a set of impressions
$V$ that arrive online. We also have a set of {\em dimensions} $K_u$ 
for each bidder $u$ that 
represent the smallest level of user-attribute granularity over which the bidder defines her budgets. 
In our previous example, one dimension 
would correspond to ``Californian females living in Los Angeles in the age range 20-29''; 
another dimension would be ``Californian females living outside of 
Los Angeles in age range 40+''.  For simplicity of notation, we will consider a universal
set of dimensions $K = \cup_u K_u$ and assume that each bidder has this same set of dimensions;
dimensions in $K \setminus K_u$ will simply earn no revenue and have no budget constraints
for bidder $u$.
If an impression $v$ is assigned to an advertiser $u$, then the algorithm 
earns $r_{uv}^{(k)}$ (called the {\em bid value} in dimension $k$)
on each dimension $k\in K$. 
Note that we allow multiple non-zero $r_{uv}^{(k)}$ revenue entries for a single 
impression-bidder pair $(u,v)$. This is to reflect the fact that certain categories are
more definite than others. For instance, the campaign that an advertisement belongs to 
is known but attributes of an online user may be less certain. If an anonymous online user 
has a 60\% chance of being below 30 and 40\% of being above 30 (e.g., based on browsing behavior), 
then the revenue earned for this user should be split in the same proportion 
between these two user category dimensions. In this case, the bid vector has a non-zero entry 
in multiple dimensions. (The reader is referred to Chatwin~\cite{Chatwin13} for 
an overview on uncertainty in determining online user attributes.
Adhikari and Dutta discuss how attribute uncertainty is weighted in real-time bidding 
strategies~\cite{AdikariD15}, while Ghosh {\em et al.}~\cite{GhoshMPV09} highlights the 
different levels of information about the users available to different entities in Internet 
advertising. Techniques for determining user attributes based on historical and prior behavior
have also been extensively studied in the marketing research community
(see Barajas Jamora~\cite{Barajas15} and references contained therein), where the mapping of 
user behavior to attributes is inherently probabilistic.) 

The revenue generated from an advertiser $u$ is subject to an arbitrary set of budget
constraints $S_u$, where each constraint $s \in S_u$ caps the total revenue generated from a 
subset of dimensions $K_s$ to a budget $B_u^{(s)}$ (see Figure \ref{fig:probdef} to see how
subsets of dimensions are used to define the budgets in our example). 
As usual, the objective in the \gen problem is to maximize the total 
revenue generated by the algorithm.  We will measure our algorithms using 
{\em competitive analysis}, which is the maximum ratio over all instances of the objectives of
an optimal solution and the algorithmic solution (see, e.g.,~\cite{BorodinE98}).

A natural question is whether this generalization changes the structure of the 
\adwords problem. To understand
this, let us consider an instance with a single advertiser. In this
case, an algorithm that assigns all impressions to the lone advertiser is 
clearly optimal for the \adwords problem. However, let us now consider the 
\gen problem with 2 dimensions, and budget constraints of \$1 each on dimensions $\{1, 2\}$
and $\{2, 3\}$. Now, suppose the first impression has a revenue of \$1 on 
dimension 2 alone. Should the algorithm assign the impression to the lone advertiser?
If it does, then the instance will generate two impressions yielding a revenue of 
\$1 on dimensions 1 and 3 each, while if it does not, then the instance will 
generate no other impression. Clearly, this example shows that no algorithm can 
do better than a competitive ratio of 2, even with a single advertiser. One 
may object that the small bids assumption is being violated, but replacing an 
impression of bid value \$1 with $1/\epsilon$ impressions of bid value 
$\epsilon \rightarrow 0$ still produces a constant lower bound of 3/2. 
In fact, this lower bound is
a manifestation of a more general observation: {\em it may
be a better option for the algorithm to not allocate an impression, or to not
earn revenue from some of the dimensions, even when possible to do so}. This is 
in sharp contrast to the classical \adwords problem, where an impression
should always be allocated if possible. Thus, the \gen problem introduces an 
aspect of ``admission control'' to the \adwords framework.
Our results for \gen are characterized in the following theorem. 
\begin{theorem} 
\label{thm:gen}
The competitive ratio of the \gen problem is $\Theta(\lg p)$ under the small bids assumption, where $p = \max_{u\in U, k\in K} |\{s: k\in K_s\}|$ 
	denotes the the maximum number of budget constraints for an advertiser to which any dimension 
	belongs.
\end{theorem} 


Although there is a super-constant lower bound for \gen, we observe that many multi-tier budgets 
based on real-world instances will have additional structure. As we saw earlier in our motivating 
example, each budget constraint was a subdivision of a more general constraint, i.e., 
the total budget for Californian females was divided into age groups to obtain the second level of budgets, 
and then each of these age groups was divided based on residency to obtain the next level. 
In essence, many consumer and product classifications are naturally hierarchical. If 
the budgets of an \gen instance are defined over such a taxonomy, 
then we have the additional structure that the budget sets $\{K_s : s \in S_u\}$ for each bidder $u$ form a 
 {\em laminar} family, i.e., for every pair 
of intersecting sets in $\{K_s : s \in S_u\}$, one is contained in the other. 

  


Thus, we also consider the
{\em \adwords  with laminar budgets} problem (\lam).
It turns out that laminar budgets make admission control redundant --- the algorithm 
can now earn revenue whenever possible. However, there are other conceptual 
difficulties. Consider an instance with 2 dimensions,
where an advertiser has a budget of \$1 for dimension 1 and an overall 
budget of \$2 for dimensions $\{1, 2\}$. At any point in the algorithm, 
what is the total budget of dimension 2, i.e., cap on total revenue earned from dimension 2?
This value clearly depends on the revenue earned on dimension 1,
and therefore changes during the course of the algorithm. This is 
in sharp contrast to the classical \adwords setting where the total budget
of a bidder remains unchanged during the course of the algorithm
(note the distinction between total budget and remaining budget). 
The first technical hurdle, therefore, is to define a dynamic notion
of total budget on individual dimensions. In addition, we also need to
define a notion of current budget utilization for individual dimensions
to determine which dimensions we should prefer in making the allocation.
Again, this notion is canonical in the classical \adwords problem -- it 
is simply the fraction of the budget of a bidder that has already been 
earned as revenue. In our more general setting, a single dimension might
be in multiple budget constraints, and therefore, we must first
identify the most constraining budget. Once we do so, we need a 
mechanism for importing the budget utilization of this constraining 
budget to the dimension itself. For instance, in the example above,
if the revenue earned on dimension 1 is \$1 and that on dimension 
2 is \$0 at some point in the algorithm, then should be budget 
utilization for dimension 2 be 0.5 (from its most constraining 
budget constraint) or 0 (from the fact that the algorithm has not
earned any revenue at all from dimension 2 yet)?
It turns out that these concepts
(revenue cap, most constraining budget, and budget utilization of a dimension)
are closely tied to each other and have to be defined through a
common inductive process. Our main technical contribution 
for the \lam problem is to carefully define these entities 
in a way that ensures semantic consistency and eventually 
%
gives our main result for this problem:
an algorithm with a competitive ratio of
$e/(e-1)$, matching that for the classic {\sc adwords} problem \cite{MehtaSVV07}. 


 \begin{theorem}
\label{thm:lam}
The competitive ratio of \lam is $e/(e-1)$ under the small bids assumption. 
\end{theorem}

Finally, we study the \gen problem {\em without} the small bids assumption. 
In the absence of this assumption, there are two possible variants -- either 
(a) the algorithm can choose the amount of revenue it earns on any given dimension
(which can be less than the corresponding bid value) from an impression, or (b) the algorithm is 
constrained to earn the entire bid value as revenue on any dimension, which 
means that an assignment of an impression to an advertiser is only allowed if 
adding the bid value to the previously earned revenue on each dimension does
not violate any constraint. The former is more natural in the context of 
Internet advertising -- we call it the \genp problem\footnote{P for ``partial''} 
and match the bounds in the small bids case.

\begin{theorem}
\label{thm:genp}
  The competitive ratio of the \genp problem is $\Theta(\lg p)$, where $p = \max_{u\in U, k\in K} |\{s: k\in K_s\}|$ 
	denotes the the maximum number of budget constraints for an advertiser to which any dimension 
	belongs. 
\end{theorem}

We also study the latter problem (where the entire bid value is always 
added to the revenue), primarily because of interesting connections 
to the classical online admission control problem~\cite{AwerbuchAP93}.
In Theorem~\ref{thm:genaon}, 
we give the competitive ratio of the
\genaon problem (AON for ``all or nothing"). As a byproduct of our result, we also obtain tight
bounds for the online admission control problem, slightly improving 
the classical bounds of \cite{AwerbuchAP93}. 
%
%
%
\begin{theorem}
\label{thm:genaon}
	Let $\epsilon = \max_{u, v, s} \frac{\sum_{k\in K_s} r_{uv}^{(k)}}{B_u^{(s)}}$ 
	denote the maximum bid-to-budget ratio and $p = \max_{u\in U, k\in K} |\{s: k\in K_s\}|$ 
	denote the the maximum number of budget constraints for an advertiser that a dimension 
	belongs to. If $ \frac{1}{\lg (2p)} < \epsilon < 1$, then the competitive ratio  of the \genaon problem is  $\Theta\left(\frac{p^{\frac{\epsilon}{1-\epsilon}}}{\epsilon}\right)$.
\end{theorem}
Note that we do not consider $\epsilon \leq \frac{1}{\lg (2p)}$ in Theorem \ref{thm:genaon}, as in this case \genaon is essentially 
identical to \gen with small bids.

 For reasons of 
brevity, we do not discuss the admission control problem here -- the 
implication of the above theorem to this problem is 
straightforward -- and relegate the details for the large bids 
case, both \genp and \genaon, to the appendix.


%
%
%
\medskip
\noindent
{\bf Related Work.} 
Given the large volume of work in this area, we will only sample 
a small fraction of the online matching and \adwords literature, 
focusing on results in the (adversarial) online model. For a 
comprehensive survey, including results in stochastic input models,
the reader is referred to the survey by Mehta~\cite{Mehta13}. 

Karp, Vazirani, and Vazirani~\cite{KarpVV90} introduced the online 
matching problem, and gave a tight $e/(e-1)$-competitive algorithm
(see also \cite{GoelM08}, \cite{BirnbaumM08}, and \cite{DevanurJK13}).
The first generalization was to the $b$-matching problem by
Kalyanasundaram and Pruhs~\cite{KalyanasundaramP00}. Later 
generalizations include a vertex-weighted version
by Aggarwal~{\em et al.}~\cite{AggarwalGKM11}, 
a {\em pay-per-click} model using stochastic rewards
by Mehta and Panigrahi~\cite{MehtaP12} (see also \cite{MehtaWZ15}),
and a bi-objective model suggested by Aggarwal~{\em et al.} \cite{AggarwalCMP14}.
Devanur and Jain~\cite{DevanurJ12} explored non-linear concave objectives 
to encode, e.g., penalties for under-delivery.
In terms of techniques, most of the initial results used combinatorial methods,
but recent work has focused on a (randomized)
primal dual technique introduced by Devanur {\em et al.} \cite{DevanurJK13}.

The \adwords problem, which generalizes $b$-matching, 
was introduced by Mehta~{\em et al.} \cite{MehtaSVV07},
who gave an $e/(e-1)$ approximation for small bids. They also showed that 
this competitive ratio is the best possible.
Without the small bids assumption, the greedy algorithm for the 
\adwords problem has a competitive ratio of $2$, and while this is 
tight for deterministic algorithms, obtaining a better ratio using a
randomized algorithm is open. 
Buchbinder~{\em et al.}\ \cite{BuchbinderJN07} gave an alternative primal-dual
analysis for the algorithm of Mehta~{\em et al.} \cite{MehtaSVV07}
with the same competitive ratio. More recently, other variants of the \adwords problem 
have been considered.  For instance, Feldman~{\em et al.}~\cite{FeldmanKMMP09} and
Aggarwal {\em et al.}~\cite{AggarwalGKM11} introduced variants to model
display ads with vertex weights and/or capacities. \smallskip

%

\noindent {\bf Paper Organization.} In Section \ref{sec:lam-sb}, we prove an $e/(e-1)$ upper bound for \lam under the small bids assumption (Theorem \ref{thm:lam}). 
In Section \ref{sec:gen}, we prove an $O(\lg p)$ upper bound for \gen under the small bids assumptions (Theorem \ref{thm:gen}). Our results for \gen without the
small bids assumptions (Theorems \ref{thm:genp} and \ref{thm:genaon} for \genp and \genaon, respectively), as well as our $\Omega(\lg p)$ lower bound for \gen, can be found in the appendix.

%% file: ec-mm-laminar-small.tex

%
%

\subsection{Primal and Dual Formulations}
\label{subsec:primaldual}
Our algorithm uses a primal-dual formulation of the \lam problem. In other words,
we give a primal LP formulation of the \lam problem and its corresponding dual, 
and update the solutions to both LPs on the arrival of a new impression. The 
primal updates, which are guided by the dual solution, define the algorithm.
The main challenge is to show that the dual updates maintain feasibility 
while ensuring that the ratio of the primal and dual objectives remains 
bounded by the desired competitive ratio $\rho = e/(e-1)$.

%
%
Let us define the {\em current budget utilization} for constraint $s$ of bidder $u$
(denoted $\kappa_u^{(s)}$) to be the fraction of budget $B_u^{(s)}$ currently 
used by the algorithm, or formally, $\kappa_u^{(s)} = \sum_{k \in K_s, v} r_{uv}^{(k)}/B_u^{(s)}$ 
for impressions $v$ assigned to bidder $u$ thus far.
Let us call a dimension $k$ {\em active} for bidder $u$ if for all budgets 
$s \in S_{u}$ such that $k \in K_s$, the algorithm currently has $\kappa_u^{(s)} < 1$. 
In other words, $u$'s active dimensions are the ones on which the algorithm can still receive 
revenue from $u$. 

An algorithm for \lam might gain revenue from only a subset of dimensions when 
assigning an impression to a bidder. To implement this flexibility in the LP, 
we introduce the notion of {\em assignment types}. 
Let $t \subseteq K$.  For impression $v$ and bidder $u$, we define a {\em type-$t$ assignment} as 
one where the dimensions in $t$ are active and the dimensions in $K\setminus t$ are inactive. 
Thus, our decision variables for the LP will be of the form $x_{uvt}$, where the algorithm sets $x_{uvt}$ to be 1 if impression 
$v$ is assigned to bidder $u$ using a type-$t$ assignment (and 0 otherwise). We then define $r_{uvt}^{(k)} = r_{uv}^{(k)}$ if $k \in t$; otherwise,  $r_{uvt}^{(k)} = 0$. 
Our primal LP $P$ is now defined as:
\begin{gather} 
\max \sum_{u,v,t} x_{uvt} \sum_{k} r_{uvt}^{(k)} \notag \\ 
\forall \ u \in U,~ s \in S_u: \quad  \ \sum_{v,t} x_{uvt} \sum_{k \in K_s} r_{uvt}^{(k)} \leq B_{u}^{(s)} \label{eq:primal1} \\
\forall \ v \in V : \quad  \sum_{u,t} x_{uvt} \leq 1 \label{eq:primal2} \\
\forall \ u \in U,~ v \in V,~ t \subseteq K: \quad  x_{uvt} \geq 0 \notag . 
\end{gather} 
Note that Eq.~\eqref{eq:primal1} ensures that the algorithm receives
no revenue from inactive dimensions for a bidder, and Eq.~\eqref{eq:primal2}
ensures that every impression $v$ is assigned using a single type to a single bidder.

The dual $D$ of this LP is defined as: 
\begin{gather}
\min \sum_u \sum_{s \in S_u} \alpha_{u}^{(s)} B_{u}^{(s)} + \sum_v \sigma_v \notag \\ 
\forall \ u \in U,~ v \in V,~  t \subseteq K : \quad  \sum_{s \in S_u} \left(\alpha_{u}^{(s)} \sum_{k \in K_s} r_{uvt}^{(k)}\right)  + \sigma_v \geq \sum_{k} r_{uvt}^{(k)} \label{eq:dual-basic-1} \\
\forall \ u \in U,~ s\in S_u: \quad  \alpha_{u}^{(s)} \geq 0 \notag\\ 
\forall \ v \in V: \quad \sigma_v\geq 0. \notag
\end{gather} 

Unfortunately, the dual stated above cannot be used directly in a primal dual algorithm.
If a constraint $s$ has budget utilization $\kappa_u^{(s)} = 1$, then the 
dual variable $\alpha_u^{(s)}$ also needs to be equal to 1 in order to balance the 
contributions to the two sides of the dual constraint by dimensions $k\in K_s$.
(Note that the primal objective does not increase for these dimensions and hence 
the value of $\sigma_v$ in the dual objective cannot depend on these dimensions 
either, if the ratio of the primal to dual objective is to be maintained.)

Now, if we na\"ively enforce $\alpha_u^{(s)} =1$ once a budget $B_u^{(s)}$ is tight, then the 
primal-dual ratio could be proportional to the number of nested levels, if a nested set of budgets are
all tight. To obtain a constant competitive ratio, what our scheme will (roughly speaking) need to do 
is only set $\alpha_u^{(s)}$ to 1 at the highest level of nesting for each nested set of tight constraint. 
If we think of the primal objective as being ``attributed'' to dual variables in order to maintain the 
primal dual ratio, then what we roughly want is that at any point of time, the primal objective from a 
given dimension for some bidder $u$ is attributed to a unique dual variable representing a budget 
constraint for $u$ containing that dimension.
However, in order to implement this property in an online setting, $\alpha_u^{(s)}$ variables need to be 
non-monotone since the budgets in lower nesting levels might become tight first followed by the higher 
levels.  Therefore, we need a 
means of raising and lowering each $\alpha_u^{(s)}$ so that at the end of the 
instance, the revenue earned from a dimension for bidder $u$ is attributed 
to exactly one of these variables.  In general, non-monotonicity of dual variables is 
undesirable in online algorithms because a satisfied dual constraint might become unsatisfied 
later. To overcome this problem, we give a new dual $D'$ where we decompose $\alpha_u^{(s)}$ 
into decision variables that are indeed monotone in our eventual primal-dual analysis. 


Formally, our transformed dual $D'$ is defined as follows. 
Since each $S_u$ is laminar,
we can represent its set system as a forest $F_u$, where each node 
in the forest corresponds to a constraint $s \in S_u$, and the 
singleton budgets are the leaves.  
%
Let $A_s$ be the set of ancestors of $s$ in $F_u$,
including $s$ itself.  
Define a new decision variable 
$\gamma_u^{(s)} = \sum_{s' \in A_s} \alpha_{u}^{(s')}$,
and let  $p(s)$ be the parent budget of $s$. 
Observe that $\alpha_{u}^{(s)} = \gamma_u^{(s)} - \gamma_u^{(p(s))}$ (where 
for a maximal set $s$ with no parent in $F_u$, we set $\gamma_u^{(p(s))} = 0$).
Using the new variables, we can rewrite Eqn.~\eqref{eq:dual-basic-1} in our original dual formulation as:
\begin{gather}
\forall \ u \in U,~ v \in V,  t \subseteq K : \quad \sum_{s \in S_u} \left(\left(\gamma_u^{(s)} - \gamma_u^{(p(s))}\right) 
                                                        \sum_{k \in K_s} r_{uvt}^{(k)}\right)  + \sigma_v \geq \sum_{k} r_{uvt}^{(k)}. \label{eq:dualre2}
\end{gather} 
Next, we observe that the outermost summation on the LHS of Eqn.~\eqref{eq:dualre2} telescopes, 
and the only remaining $\gamma_{u}^{(s)}$ are those for singleton budgets. 
This gives us our final dual formulation $D'$:


 \begin{gather}
\min \sum_v \sigma_v +\sum_u \sum_{s \in S_u} B_u^{(s)} \left(\gamma_{u}^{(s)} - \gamma_u^{p(s)}\right) \notag \\ 
\forall \ u \in U,~ v \in V,  t \subseteq K : \sum_{k}  \gamma_{u}^{(\{k\})} r_{uvt}^{(k)}  + \sigma_v \geq \sum_{k} r_{uvt}^{(k)} \label{eq:dual1} \\
\forall \ u \in U,~ s\in S_u: \quad \gamma_u^{(s)} - \gamma_u^{(p(s))} \geq 0 \label{eq:dual2}.
\end{gather}

\noindent

%


\subsection{Labeling Scheme}  
Recalling our above discussion, our goal will be to attribute the revenue earned on a dimension $k$ for 
bidder $u$ to exactly one dual variable $\gamma_u^{(s)}$, 
ideally to the $\gamma_u^{(s)}$ corresponding to the ``most-limiting" budget $B_u^{(s)}$
such that $k \in K_s$.
This suggests that we should make $\gamma_u^{(s)}$ a monotonically increase 
function of the budget's current utilization $\kappa_u^{(s)}$.
However, simply using utilization to define $\gamma_u^{(s)}$ does not capture the interactions 
between budgets in the laminar setting. 
The overarching issue with just using $\kappa_u^{(s)}$  is the fact that $B_u^{(s)}$ might be the most utilized constraint for only {\em some} of the dimensions in $K_s$, since other budgets that sit below $B_u^{(s)}$ in the hierarchy may have
higher utilization. This raises the following question: should the revenue currently constrained by these descendant budgets, 
say revenue earned on some particular dimension $k'$, affect how the algorithm determines the extent to which
$B_u^{(s)}$ limits other unbounded dimensions like $k$? The answer is not immediate. It is tempting to say ``no'' 
since the dimension-$k'$ revenue
is already bounded by a tighter budget; on the other hand, $B_u^{(s)}$
might in fact become the tightest budget for dimension $k'$ later in the 
instance and ignoring the dimension-$k'$ revenue till that time will 
prevent a smooth transition of the tightest budget for $k'$.


To overcome this challenge, we introduce a labeling scheme $\ell_u^{(s)}: \cup_u S_u \rightarrow [0, 1]$. 
 Label $\ell_u^{(s)}$ for budget $B_u^{(s)}$ will represent the modified notion of the budget's utilization that we need to properly
measure the remaining capacity for future revenue. Our primal-dual analysis will then follow by making each dual variable
a monotone function of these labels. 

More concretely, we address the above challenge by having our labels maintain the following two high-level features: 
\begin{itemize} 
\item For label $\ell_u^{(s)}$ and bidder $u$, revenue from a dimension $k \in s$ will only contribute to the label if $\ell_u^{(s)}$ is at least as large as the labels of all budgets containing $k$ that are subsets of $s$.  This corresponds to identifying the ``most constrained'' budget for any dimension by interpreting these abstract labels as surrogates of actual budget utilizations.
\eat{
In other words, if there is a budget $B_u^{(s')}$ along this path that is more utilized than $B_u^{(s)}$ 
(i.e., has a higher label),
we will think of $B_u^{(s')}$ as the budget that currently bounds the revenue for dimension $k$ (w.r.t.\ to the budgets in the subtree rooted at $s$), and therefore  
dimension-$k$ revenue will not count towards measuring the utilization of budget $B_u^{(s)}$.  With regards to our earlier question, the reason we err on the side of {\em not} including 
the bounded revenue is to ensure that we do not overcharge the dual. Again recall that we want attribute the current change in primal cost earned on a dimension $k$ to exactly one dual variable $\gamma_u^{(s)}$, and such a scheme will just simply be infeasible if dimension-$k$ revenue counts towards the utilization of  labels corresponding to slack budgets.
}
%
%
%
%
\item 
In defining label $\ell_u^{(s)}$, we need to identify the capacity of constraint $s$ for future revenue earnings
from the dimensions that are deriving their label from $s$. We define this capacity as the total budget $B_u^{(s)}$ 
minus the budgets of constraints below $s$ that have a higher label. This automatically discounts the revenue earning 
capacities of dimensions that are deriving labels from descendant constraints of $s$.
\eat{
Given we enforce this first property and err towards excluding bounded revenue from the labels, we still need a means of encoding how this discounted revenue affects the utilization. 

Our solution will be to carefully lower (and then potentially raise) the {\em capacity size} of the labels. 
}
\end{itemize} 

One challenge with maintaining these properties is that they are somewhat circular. 
To determine the value of a label, we need to first determine which dimensions count toward the label, 
but determining dimension inclusion requires comparisons between label values. 
 Another challenge is maintaining smoothness. As impressions are assigned to bidders, budgets that were previously slack will  become tight, which requires us to 
 reassign dimensions to labels and change their capacities. In order to make our primal-dual analysis smooth, we will need to ensure that labels remain consistent 
 after we reassign dimensions to labels and change label capacities.
 
To overcome the issue of circularity, we will not give an explicit label definition but rather give a set of label properties that we maintain throughout the algorithm. 
These properties are based on two sets of budgets, $L(s)$ and $T(s)$, that the algorithm will dynamically update for all bidders $u$ and budgets $B_u^{(s)}$ (we drop the subscript $u$ 
for simplicity).  
The two sets partition the descendant dimensions of $s$ (i.e., each descendant dimension belongs to exactly one set in $L(s) \cup T(s)$). 
Intuitively, $L(s)$ contains singleton budgets $\{k\}$ representing
dimensions that count toward label $\ell_u^{(s)}$. On the other hand, 
$T(s)$ contains the closest descendants of $s$ that have a bigger label than $s$, i.e., every dimension in $s$ that is not in $L(s)$ derives its
label from a budget in $T(s)$ or from one their respective descendants.



Let $R_u^{(k)}$ be the total revenue currently earned on dimension $k$ for bidder $u$. 
Formalizing the above discussion, we say the labels for bidder $u$ are {\em valid} if the following three properties hold for all $s \in S_u$. 

\begin{enumerate} 

\item  {\em Property 1:} For all $\{k\} \in L(s)$,  all constraints $s'$ on the path from $\{k\}$ 
	to $s$ in $F_u$ have $\ell_u^{(s')} \leq \ell_u^{(s)}$. 
	
\item {\em Property 2:} For all $s' \in T(s)$, we have $\ell_u^{(s')} > \ell_u^{(s)}$, and for all $s''$ on the path from $s'$ to $s$, we have 
	$\ell_u^{(s'')} \leq \ell_u^{(s)}$.
	
\item {\em Property 3:}   The following identity holds: 
        \begin{equation} 
	\ell_u^{(s)} = \frac{\sum_{\{k\} \in L(s)} R_u^{(k)}}{B_u^{(s)}  - \sum_{s' \in T(s)} B_u^{(s')}}. \label{eq:prop3} 
	\end{equation} 

\end{enumerate} 
Note that once we have fixed sets $T(s)$ and $L(s)$ for all $s$, we can verify all three properties and use Eqn.\ \eqref{eq:prop3}
to directly compute each label for each $s \in S_u$. Also observe we can initialize all labels to be 0, and  set $T(s) = \emptyset$ and 
$L(s) = \{\{k\} : k \in K\}$ to start with a valid labeling.  
Finally, we note that we will soon show that all labels remain non-negative
(in the proof of Lemma~\ref{lma:label-monotone}).

\begin{figure} 
\begin{center}
 \hspace{0in}
\resizebox{4.3in}{!}{\includegraphics{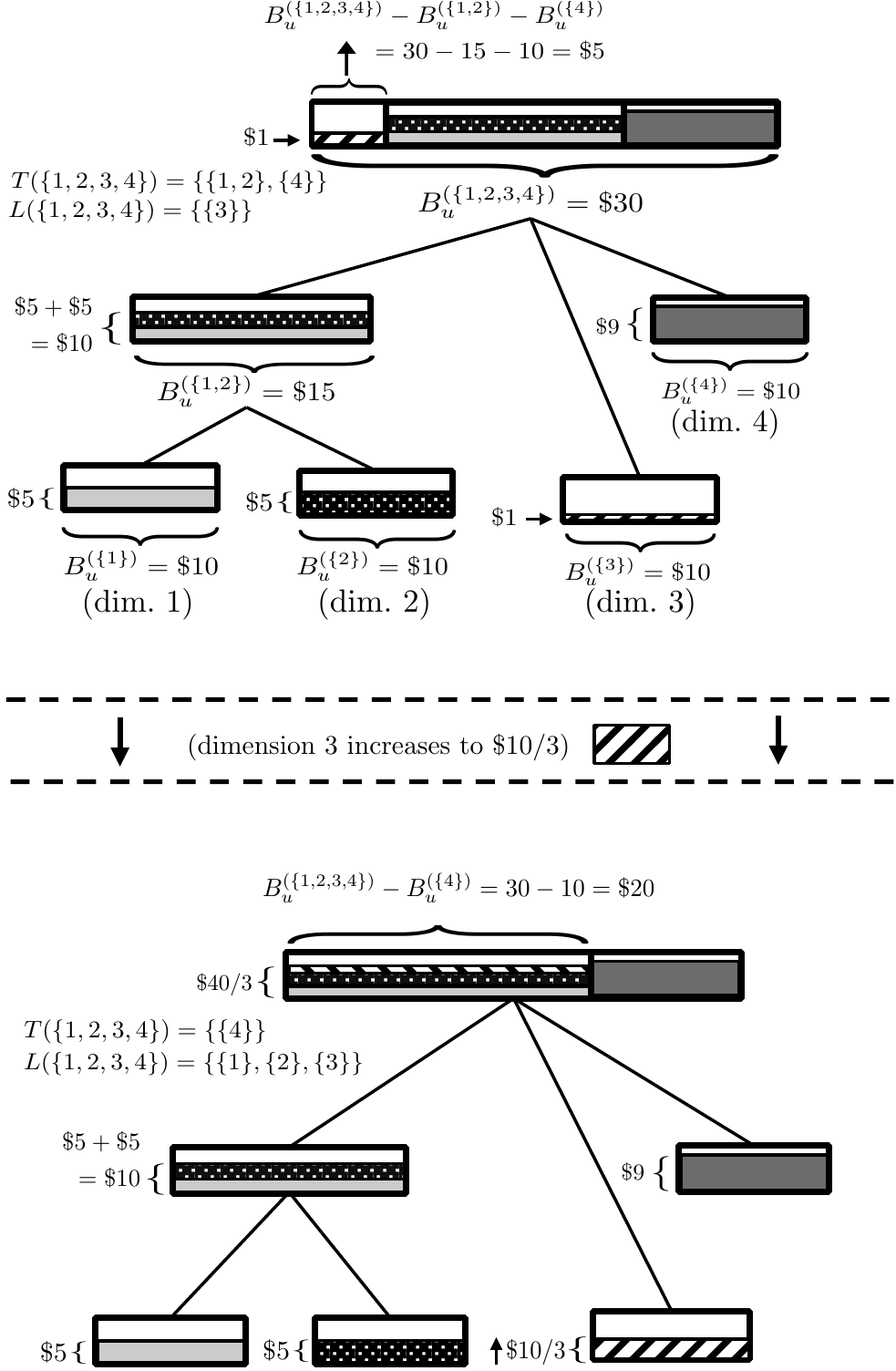}} 
\vspace{.3in}
\caption{{\small  Illustration of the modifications made to $L(s)$ and $T(s)$ in Event 2, occurring between budget sets $s = \{1,2,3,4\}$ and 
$s' = \{1,2\}$. In this example, the algorithm is incrementing the revenue on dimension 3. 
The state of the labels before dimension 3 revenue is added is shown on top.  Notice that $\{3\} \in L(\{1,2,3,4\})$ since 
$\ell_u^{\{3\}} = 1/10$ is smaller than $\ell_u^{\{1,2,3,4\}} = 1/5$  (note that the value of $\ell_u^{\{1,2,3,4\}}$ corresponds to the left most rectangle, where 
the middle and right rectangles show the revenue levels of the dimensions that are {\em not} included in the label). After increasing dimension 3 to \$10/3, we have that
$\ell_u^{\{1,2,3,4\}} = 2/3$, which means it is about to surpass label $\ell_u^{\{1,2\}}$, therefore triggering Event 2. The state of the labels after the Event 2 modifications
 are shown on bottom (at the transition point). Notice that $\{1,2\}$ has 
 been removed from $T(\{1,2,3,4\})$, and we have added the sets in $L(\{1,2\})$ to $L(\{1,2,3,4\})$ (namely, $\{1\}$ and $\{2\}$).}} 
\label{fig:event2} 
\end{center}
\end{figure}

Next, we define a procedure for updating labels after the revenue earned in a single dimension increases
infinitesimally. More specifically, suppose an impression $v$ 
is assigned to bidder $u$, and assume we have a valid labeling for the sets in $S_u$ before the assignment.   
The assignment of the impression changes the values of $R_u^{(k)}$ for the active dimensions $k$. This necessitates label updates, which we define for 
an infinitesimally small increment in the value of $R_u^{(k)}$ for a particular dimension $k$.  Note that the overall assignment of the impression is a sequence of such incremental changes. 


The labels that we increase on such an increment are $\ell_u^{(s)}$
such that $\{k\} \in L(s)$, i.e., we will increase $R_u^{(k)}$ in the numerator of Eqn.\ \eqref{eq:prop3} for all such $\ell_u^{(s)}$ and leave $T(s)$ and $L(s)$ fixed.
If the labeling is valid before a given increment, and if after the increment
the relative ordering of all $\ell_u^{(s)}$ remains the same, then by the definition of Properties 1 and 2 the labeling remains valid. Thus, in terms of updating 
$L(s)$ and $T(s)$, we only need to consider when the relative order of labels changes as a result of adding to $R_u^{(k)}$. In particular, there are two types of reordering that need considered. We will call these two reordering possibilities 
Events 1 and 2 and describe how the algorithm updates $L(s)$ and $T(s)$ in each case. 
Later, we will show that these set redefinitions maintain the current value of the label. 

\begin{itemize} 
\item {\bf Event 1:} For some $s$ such that $\{k\} \in L(s)$, there now exists a descendant $s'$ of $s$ such that $\ell_u^{(s')} > \ell_u^{(s)}$, where $s'$ is on the path from $\{k\}$ to $s$
and $\ell_u^{(s')} \leq \ell_u^{(s)}$ previously. 
In this event, $s'$ is now added to $T(s)$. All singleton-budgets $\{k\}$ that are descendants of $s'$ and
belong to $L(s)$ are now removed from $L(s)$. Additionally, all descendants of $s'$ that belong to $T(s)$ are also removed from $T(s)$.

\item {\bf Event 2:} For some constraint $s$ such that $\{k\} \in L(s)$, there now exists a descendant $s'$ of $s$ such that $\ell_u^{(s)} \geq \ell_u^{(s')}$, where $s' \in T(s)$ previously. 
In this event, $s'$ is removed from $T(s)$.  Conversely to Event 1, all constraints in $T(s')$ are added to $T(s)$, and all $\{k\} \in L(s')$ are added to $L(s)$. 
\end{itemize} 
In order to make the process smooth, we will think of the updates in Events 1 and 2 as being done at the transition point where $\ell_u^{(s)} = \ell_u^{(s')}$. 
This completes the description of our labeling scheme and the process by which the algorithm determines them. We encourage the reader to refer to Figure \ref{fig:event2} for a small example of an Event 2 update.\footnote{In terms of how $L(s)$ and $T(s)$ are updated, Event 1 is the reverse of Event 2. So, reversing the example in the Figure \ref{fig:event2} will provide the reader with an Event 1 example.}  We now prove the following lemma, which will be useful for our primal-dual analysis. 


\begin{lemma}
	\label{lma:label-monotone}
	For every constraint $s$, the label $\ell_u^{(s)}$ is monotonically non-decreasing over the course of the algorithm.
\end{lemma}
\begin{proof}
Clearly when neither Event 1 or 2 occurs, $\ell_u^{(s)}$ can only increase (this follows directly from the definition of the update procedure). Thus, it suffices 
to show that $\ell_u^{(s)}$ does not decrease when it participates in Event 1 or 2. In particular, we will show that $\ell_u^{(s)}$ has an identical value after $L(s)$ and $T(s)$ have been modified 
in either event. We will show that this holds for Event 1, noting that the argument for Event 2 is identical. 

Suppose the updates for Event 1 occur for a constraint $s$ and a descendant $s'$, triggered by a dimension $k \in s$. The changes are: 
$s'$ is added to $T(s)$, all descendant singleton-budgets 
of $s'$ that were in $L(s)$ are removed from $L(s)$, and all descendants of $s'$ that were in $T(s)$ are removed from $T(s)$. Recall that before the event, we have that 
\begin{alignat*}{2} 
\ell_u^{(s)} &=  \frac{\sum_{\{k\} \in L(s)} R_u^{(k)}}{B_u^{(s)}  - \sum_{w \in T(s)} B_u^{(w)}} \\ 
\ell_u^{(s')} &= \frac{\sum_{\{k\} \in L(s')} R_u^{(k)}}{B_u^{(s')}  - \sum_{w' \in T(s')} B_u^{(w')}}. 
\end{alignat*} 

Therefore, using the definitions of $L(s)$ and $T(s)$ before they are modified by the event, the new $\ell_u^{(s)}$ (denoted $\ell_{\text{new}}^{(s)}$)   can be written as: 

\begin{equation} 
\ell_{\text{new}}^{(s)} = \frac{\sum_{\{k\} \in L(s)} R_u^{(k)} - \sum_{\{k\} \in L(s')} R_u^{(k)}}{B_u^{(s)}  - \sum_{w \in T(s)} B_u^{(w)} - B_u^{(s')} + \sum_{w' \in T(s')} B_u^{(w')}}. 
\label{eq:event1eq}
\end{equation} 
Since $\ell_u^{(s)} = \ell_u^{(s')}$ at the moment Event 1 occurs, we have $\ell_{\text{new}}^{(s)} = \ell_u^{(s)} = \ell_u^{(s')}$ (which follows from the fact that $ a/b = c/d = \alpha$ implies $(a-c)/(b-d) = \alpha$).

To complete the proof, note that the above argument does not exclude the possibility of $\sum_{\{k\} \in L(s)} R_u^{(k)}= 0$ 
and the denominator in Eqn.\ \eqref{eq:event1eq} being negative (if this were to happen, the increments to $R_u^{(k)}$ would decrease the label by making it more negative). However, in both events this cannot be the case. First observe that Event 2 can only occur between two non-zero labels (since in Event 2, $s' \in T(s)$ before the event, which implies a strict inequality $\ell_u^{(s)} < \ell_u^{(s')}$).  Event 1 can (and will) occur when $\ell_u^{(s)} = \ell_u^{(s')} = 0$, but in Event 1, the denominator 
 of $\ell_u^{(s')}$ must always be smaller than the denominator of $\ell_u^{(s)}$. This is because 
Event 1 can only occur between two labels such that $\{k\} \in L(s)$ where $k$ the dimension is currently being incremented. 
Since $\ell_u^{(s')}$ is surpassing $\ell_u^{(s)}$ at the transition point in Event 1, it must be the case that $\ell_u^{(s')}$ is increasing at a higher rate than $\ell_u^{(s)}$. This implies 
$\ell_u^{(s')}$ must have a smaller denominator than $\ell_u^{(s)}$ before the modifications to $L(s)$ and $T(s)$. \end{proof}

\subsection{Algorithm Definition and Analysis} 
Using our dual formulation and labeling scheme, we are now ready to define and analyze our algorithm. Consider the arrival of impression $v$. 
Define $g_u{(s)} = \max_{s'\in A_s} \ell_u^{(s')}$, i.e., the maximum label of an ancestor of $s$ in the forest $F_u$ (including $s$ itself). Our 
algorithm assigns impression $v$ to bidder $u = \arg\max_{u' \in U} \{D_{u'v}\}$,
where $D_{uv} = \sum_{k \in t_u} (1- e^{g_u^{(\{k\})} -1})r_{uv}^{(k)}$ and $t_u$ is the current active dimensions for bidder $u$. 


For the rest of the section, 
let $\rho = e/(e-1)$. For a primal assignment of impression $v$ to bidder
$u$, we change the dual solution by setting $\sigma_v = \rho\cdot D_{uv}$
and update $\gamma_{u}^{(s)}$ to be
\begin{equation*} 
\gamma_{u}^{(s)} = \frac{e^{g_u^{(s)}} -1}{e-1} = \rho(e^{g_u^{(s)}-1} - e^{-1}), 
\end{equation*} 
where $g_u^{(s)}$ is computed {\em after} the assignment of the current impression $v$. 

%
%
For the competitive analysis, it suffices to show that 
a) the ratio between dual and primal objectives is at most $\rho$, and 
b) the dual solution is feasible. 

\medskip
\noindent 
{\bf Primal-Dual Ratio.} Our goal is to show that when an impression $v$ is 
assigned to a bidder $u$, the change in dual objective is at most $\rho = e/(e-1)$ times that of the primal
objective. First, note that the dual objective is a function of the labels, and we have argued above that the 
labels do not change when either of Event 1 or 2 happens. Therefore, we only need to account for the change 
in the dual objective when the labels change but neither of the two events happen. Let us define $S^*_u$
as the subset of constraints in $S_u$ where the value of $\gamma_u^{(s)}$ is different from $\gamma_u^{(p(s))}$:
$$S^*_u = \{s\in S_u: \gamma_u^{(s)}\not= \gamma_u^{(p(s))}\}.$$
We can rewrite the dual objective as $\sum_{v\in V} \sigma_v + \sum_u \sum_{s\in S^*_u} \left(\gamma_u^{(s)} - \gamma_u^{(p(s))}\right) B_u^{(s)}$
since for all the other terms, the value of $\gamma_u^{(s)} - \gamma_u^{(p(s))} = 0$. For any constraint
$s\in S^*_u$, let $p^*_u(s)$ be its closest ancestor in $F_u$ that is also in $S^*_u$. Now, observe  that by 
Property 2 of labels and the definitionof  $g_u^{(s)}$, $T(s) = \{s'\in S^*_u: p^*_u(s') = s\}$ for any $s\in S^*_u$.
Then, the dual objective can be further rewritten as 
$$\sum_{v\in V} \sigma_v + \sum_u \sum_{s\in S^*_u} \gamma_u^{(s)} \left(B_u^{(s)} - \sum_{s'\in T(s)} B_u^{(s')}\right).$$

As earlier, we will analyze the change in the dual and primal objectives when the revenue on 
a dimension $k$ is incremented by an infinitesimal amount $\Delta r_{uv}^{(k)}$.
 Note that for any singleton constraint $\{k\}$, there is a unique $s\in S^*_u$ satisfying $k\in L(s)$; furthermore, 
 $g_u^{(\{k\})} =  g_u^{(s)} = \ell_u^{(s)}$. Therefore, the only 
dual variable in $S^*$ (i.e., in the dual objective given above) that changes is $\gamma_u^{(s)}$. 
Let us denote the change in $g_u^{(s)}$ by $\Delta g_u^{(s)}$.
Using the small bids assumption, we can write:
\begin{eqnarray*} 
\Delta \gamma_u^{(s)} \cdot \left(B_u^{(s)} - \sum_{s'\in T(s)} B_u^{(s')}\right) 
& = & \frac{\partial \gamma_u^{(s)}}{\partial g_u^{(s)}}\cdot \Delta g_u^{(s)} \cdot \left(B_u^{(s)} - \sum_{s'\in T(s)} B_u^{(s')}\right) \\
& = & \rho \cdot (e^{g_u^{(s)} -1}) \cdot \frac{\Delta r_{uv}^{(k)}}{B_{u}^{(s)} - \sum_{s'\in T(s)} B_u^{(s')}} \cdot \left(B_u^{(s)} - \sum_{s'\in T(s)} B_u^{(s')}\right) \\
& = & \rho \cdot (e^{g_u^{(\{k\})} -1}) \cdot \Delta r_{uv}^{(k)}.
\end{eqnarray*} 
Summing over all the infinitesimal changes in revenue, the total change in the dual objective for the assignment of impression $v$ is given by
$\sigma_v + \rho \cdot \sum_{k\in t_u} (e^{g_u^{(\{k\})} -1}) \cdot r_{uv}^{(k)}$,
where $t_u$ is the set of active dimensions.
Since the algorithm sets 

\begin{equation*}
\sigma_v = \rho \cdot \sum_{k \in t_u} (1- e^{g_u^{(\{k\})} -1})r_{uv}^{(k)},
\end{equation*}
the total change in the dual objective can be written as:

\begin{equation*}
	\rho \cdot \sum_{k \in t_u} (1- e^{g_u^{(\{k\})} -1})r_{uv}^{(k)} + \rho \cdot \sum_{k\in t_u} (e^{g_u^{(\{k\})} -1}) \cdot r_{uv}^{(k)}
	= \rho \cdot \sum_{k \in t_u} r_{uv}^{(k)},
\end{equation*}
which is exactly $\rho$ times the increase in the primal objective. 

\medskip
\noindent
{\bf Dual Feasibility.}
Finally, we argue that the dual is feasible when the algorithm terminates. 

\begin{lemma} 
At the end of the algorithm, the dual is feasible. 
\end{lemma} 

\begin{proof}
%
The feasibility of Eqn.~\eqref{eq:dual1} follows directly from definition of $g_u^{(s)}$, and 
the fact $\gamma_u^{(s)}$ is a non-deceasing function of $g_u^{(s)}$. 
%
We now show Eqn.~\eqref{eq:dual2}.
Let $t_u$ be the set of active dimensions for bidder $u$ when impression $v$ arrived. First, observe that for all $k \not\in t_u$, we have 
that $g_u^{(\{k\})} = 1$. This follows from the fact that if $k$ is inactive, a constraint $s \in S_u$ containing
$k$ has reached $\kappa_{u}^{(s)} = 1$ (and thus $\ell_u^{(s)} = 1$ as well).  Since $s$ is an ancestor of $\{k\}$ in $F_u$, 
we also have $g_u^{(\{k\})} = \gamma_u^{(\{k\})} = 1$. 

Let $u'$ be the bidder to which the algorithm assigned impression $v$. Let $\widehat{g_{u}^{(s)}}$ be the value of 
$g_{u}^{(s)}$ when $v$ was assigned (and define $\widehat{g_{u'}^{(s)}}$ similarly). We have the following: 

\begin{alignat}{2} 
\sum_{k}  \gamma_{u}^{(\{k\})} r_{uvt}^{(k)}  + \sigma_v  & =  \quad \sum_{k}  \gamma_{u}^{(\{k\})} r_{uvt}^{(k)}  + \rho \sum_{k \in t_u} (1- e^{\widehat{g_{u'}^{(\{k\})}} -1})r_{u'v}^{(k)}
 \hspace{5mm} (\text{by substituting } \sigma_v) \notag \\
& = \sum_{k \not\in t_u}r_{uvt}^{(k)} +  \rho \sum_{k \in t_u} (e^{g_u^{(\{k\})} -1} - 1/e) r_{uvt}^{(k)}  + \rho \sum_{k \in t_u} (1- e^{\widehat{g_{u'}^{(\{k\})}} -1})r_{u'v}^{(k)},
 \label{eq:dualfeas1}
\end{alignat} 
where the second equality follows by substituting $ \gamma_{u}^{(\{k\})}$ and the fact that 
$\gamma_u^{(\{k\})} = 1$ for all $k \not\in t_u$.  We can now  establish Eqn.\ \eqref{eq:dual2} as follows:

\begin{alignat*}{2} 
\sum_{k}  \gamma_{u}^{(\{k\})} r_{uvt}^{(k)}  + \sigma_v  &\geq \sum_{k \not\in t_u}r_{uvt}^{(k)} +  \rho \sum_{k \in t_u} (e^{g_u^{(\{k\})} -1} - 1/e) r_{uvt}^{(k)}  + \rho \sum_{k \in t_u} (1- e^{\widehat{g_{u}^{(\{k\})}} -1})r_{u'v}^{(k)} \\
& \geq \sum_{k \not\in t_u}r_{uvt}^{(k)} +  \rho \sum_{k \in t_u} (e^{g_u^{(\{k\})} -1}- 1/e) r_{uvt}^{(k)}  + \rho \sum_{k \in t_u} (1- e^{g_{u}^{(\{k\})} -1})r_{uv}^{(k)} \\
&  \geq \sum_{k \not\in t_u}r_{uvt}^{(k)} + \rho \sum_{k \in t_u}(1 - 1/e)r_{uvt}^{(k)} =   \sum_{k}r_{uvt}^{(k)}   \hspace{5mm} \text{(since $\rho = e/(e-1)$).} 
\end{alignat*}  
The first inequality follows from Eqn.~\eqref{eq:dualfeas1} and the fact that the algorithm assigns $v$ to $u' = \arg\max_{u \in U} \{D_{uv}\}$. The second inequality is because $1- e^{g_{u}^{(\{k\})} -1}$ is a non-increasing function of 
$g_{u}^{(\{k\})}$, and the third inequality follows since $r_{uv}^{(k)} \geq r_{uvt}^{(k)}$. 
\end{proof}

%% file: ec-mm-general.tex
Recall the \gen problem: we given a set of offline bidders $U$ and a set of impressions $V$ that arrive online, 
where each bidder-impression pair $(u,v)$ is specified by a bid value $r_{uv}^{(k)}$ for each dimension $k \in K$.
The revenue generated from a bidder $u$ is subject to an arbitrary set of budget
constraints $S_u$, where each constraint $s \in S_u$ caps the total revenue generated from a 
subset of dimensions $K_s$ to a given budget $B_u^{(s)}$.

In this section, we will prove an $O(\lg p)$ upper bound for \gen (Theorem \ref{thm:gen}). 

\subsection{Algorithm Definition} 
As in Section \ref{sec:lam-sb}, let $\kappa_u^{(s)}$ denote the current utilization of budget $B_u^{(s)}$.  
The algorithm (we call it \algo) uses an exponential potential function defined by:
\begin{equation*}
	\phi = \sum_u \sum_s \phi_u^{(s)} = \sum_u \sum_s \frac{B_u^{(s)}}{p} \left((2p + 2)^{\kappa_u^{(s)}} - 1\right),
\end{equation*}
where $\kappa_u^{(s)}$ is defined as the fraction of $B_u^{(s)}$ that has already been used by the algorithm
at any stage. Note that $\phi = 0$ initially.

At any stage of \algo, a dimension $k$ is said to be 
	{\em active} for bidder $u$ if and only if $\sum_{s: k\in s} \frac{\phi_u^{(s)}}{B_u^{(s)}} \leq 1$;
	otherwise, dimension $k$ is said to be {\em inactive} for bidder $u$. 
	(Note that this is a different definition of active dimensions than what is used in Section \ref{sec:lam-sb}). 
\algo only attempts to earn revenue on active dimensions, and hence, the total revenue if impression 
$v$ is allocated to bidder $u$ is given by:
\begin{equation*}
	r_{uv} = \sum_{k\in A_u} r_{uv}^{(k)}, \text{ where } A_u \text{ is the set of current active 
		dimensions for bidder } u.
\end{equation*}
The algorithm makes a greedy assignment with respect to $r_{uv}$, i.e., it assigns impression $v$ to 
$\arg\max_{u} r_{uv}$. Note that it is possible that $A_u = \emptyset$ for all bidders $u$,
and therefore the algorithm does not assign impression $v$ to any bidder, even though there are 
dimensions and bidders where it could have earned revenue. This completes the description of our algorithm. 

\subsection{Algorithm Analysis} 
For our analysis, it will be sufficient to quantify the small bids assumption as the following property
for any impression $v$, bidder $u$, dimension $k$, and constraint $s$ such that $k\in s$:
\begin{equation}
\label{eq:small}
	\sum_{k\in s} \frac{r_{uv}^{(k)}}{B_u^{(s)}} \leq \frac{1}{\lg (2p+2)}.
\end{equation}
We first establish the feasibility of the solution, which follows almost directly from how we define active dimensions.
\begin{lemma}
\label{lma:feasible}
	If \algo assigns an impression $v$ to a bidder $u$, then it can earn revenue on all the 
	active dimensions $A_u$ of $u$ without violating any constraint.
\end{lemma}
\begin{proof}
	We need to show that for all constraints $s$ of bidder $u$,
	\begin{equation*}
		\kappa_u^{(s)} + \sum_{k\in A_u\cap s} \frac{r_{uv}^{(k)}}{B_u^{(s)}} \leq 1.
	\end{equation*}
	Suppose not. Then, for some constraint $s$,
	\begin{eqnarray*}
		\kappa_u^{(s)} + \sum_{k\in A_u\cap s} \frac{r_{uv}^{(k)}}{B_u^{(s)}} & > & 1 \\
		\text{i.e., } \kappa_u^{(s)} + \frac{1}{\lg (2p+2)} & > & 1 
			\quad \text{(by the small bids assumption Eqn.~\eqref{eq:small})}\\
		\text{i.e., } (2p + 2)^{\kappa_u^{(s)} + \frac{1}{\lg (2p+2)}} & > & 2p + 2 \\
		\text{i.e., } (2p + 2)^{\kappa_u^{(s)}} & > & p + 1 \\
		\text{i.e., } \frac{p \phi_u^{(s)}}{B_u^{(s)}} + 1 & > & p + 1 \\
		\text{i.e., } \frac{\phi_u^{(s)}}{B_u^{(s)}} & > & 1,
	\end{eqnarray*}	
	which contradicts the fact that dimension $k$ is active for bidder $u$.
\end{proof}
This lemma implies that \algo is indeed able to earn revenue on all active dimensions of a bidder $u$ when 
it assigns an impression to $u$.

Next, we will bound the total revenue of an optimal solution that we denote by \opt against the total revenue
of \algo. Let $u_{\opt}(v)$ (resp., $u_{\algo}(v)$) be the bidder 
that \opt (resp., \algo) allocates impression $v$ to. For every dimension $k$ that \opt earns revenue from, 
one of the following holds:
\begin{enumerate}
	\item {\bf Case 1:} dimension $k$ is active for bidder $u_{\opt}(v)$ in \algo when impression $v$ arrives, and \algo 
		assigns $v$ to the same bidder, i.e., $u_{\algo}(v) = u_{\opt}(v)$.
	\item {\bf Case 2:} dimension $k$ is active for bidder $u_{\opt}(v)$ in \algo when impression $v$ arrives, but \algo 
		assigns $v$ to a different bidder, i.e., $u_{\algo}(v) \not= u_{\opt}(v)$.
	\item {\bf Case 3:} dimension $k$ is inactive for bidder $u_{\opt}(v)$ in \algo when impression $v$ arrives.
\end{enumerate}
We partition the dimensions that \opt earns revenue from into active and inactive dimensions (according
to their status in \algo for bidder $u_{\opt}(v)$). For active dimensions, the next lemma gives a 
straightforward charging argument using the greediness of the choice made by \algo.

\begin{lemma}
\label{lma:active}
	For any impression $v$, the total revenue earned by \opt on the active dimensions is at most the total 
	revenue earned by \algo overall.
\end{lemma}
\begin{proof}
	For case 1 above (\opt and \algo choose the same bidder), the two algorithms earn the same revenue on
	the active dimensions. For case 2 above, the fact \algo makes a greedy choice implies that it earns
	at least as much revenue by assigning to different bidder as it would have made by assigning to $u$,
	which includes the revenue on all the active dimensions.
\end{proof}	

The more involved case is that of inactive dimensions. In this case, we use a different global charging 
argument over all dimensions, based on the potential function. In particular, we show that for a bidder 
$u$, the total revenue of \opt from inactive dimensions (recall that this only includes revenue from 
impressions that arrived {\em after} the dimension became inactive in \algo) can be charged, up to a 
logarithmic loss, to the revenue that \algo earned overall from bidder $u$.

\begin{lemma}
\label{lma:loss}
	Fix a bidder $u$.  The total revenue that \opt earns in inactive dimensions for bidder $u$ is at most 
	the final potential of bidder $u$ in \algo.
\end{lemma}	
\begin{proof}
	For any dimension $k$, let $v\in V_{u, k}$ denote the subset of impressions assigned to $u$ by \opt 
	that arrived after $k$ became an inactive dimension for bidder $u$ in \algo. We need to bound the 
	total revenue earned by \opt on dimension $k$ from impressions in $V_{u, k}$, summed over all $k$. 
	For any impression $v\in V_{u, k}$, we have:
	\begin{equation*}
		\label{eq:potbound}
		\sum_{s: k\in s} \frac{\phi_u^{(s)}}{B_u^{(s)}} > 1,
	\end{equation*}
	where $\phi_u^{(s)}$ is the final potential for constraint $s$ of bidder $u$.
	Thus,the revenue that \opt earns from impressions $v\in V_{u, k}$ on dimension $k$,
	summed over all dimensions, can be bounded as follows: 
	\begin{equation*}
		\sum_k \sum_{v\in V_{u, k}} r_{uv}^{(k)} 
		< \sum_k \sum_{v\in V_{u, k}} r_{uv}^{(k)} \sum_{s: k\in s} \frac{\phi_u^{(s)}}{B_u^{(s)}}	
		= \sum_s \phi_u^{(s)} \sum_{k\in s} \sum_{v\in V_{u, k}} \frac{r_{uv}^{(k)}}{B_u^{(s)}} 
		\leq  \sum_s \phi_u^{(s)},
	\end{equation*}	
where the last inequality follows from the feasibility of \opt. 
%
%
%
\end{proof}

Finally, we need to lower bound the final potential of a bidder in terms in terms of the revenue it generates for \algo. This is done in the following lemma.
\begin{lemma}
\label{lma:pot}
	The increase in potential of a bidder $u$ during the course of $\algo$ is at most $4\lg (2p+2)$ 
	times the revenue that \algo earns from $u$.
\end{lemma}
\begin{proof}
	Suppose \algo assigns impression $v$ to bidder $u$. Let $K_a$ denote the set of active dimensions
	for bidder $u$ when this assignment is made. Let $\eta_u^{(s)} = \sum_{k: k\in K_a\cap s} \frac{r_{uv}^{(k)}}{B_u^{(s)}}$.  Then, the increase in potential is given by:
	\begin{alignat}{2} 
       \sum_s \Delta \phi_u^{(s)} 
       & = \sum_s \frac{B_u^{(s)}}{p} \left((2p+2)^{\kappa_u^{(s)} + \eta_u^{(s)}} - (2p+2)^{\kappa_u^{(s)}}\right) \notag\\
	 & = \sum_s \frac{B_u^{(s)}}{p} (2p+2)^{\kappa_u^{(s)}} \left((2p+2)^{\eta_u^{(s)}} - 1\right)  \notag \\
	 & = \sum_s \left(\phi_u^{(s)} + \frac{B_u^{(s)}}{p}\right) \cdot \left(2^{\eta_u^{(s)} \cdot \lg(2p+2)}- 1\right) \notag \\
	& \leq \sum_s \left(\phi_u^{(s)} + \frac{B_u^{(s)}}{p}\right) \cdot 2\lg(2p+2)
	   \cdot \sum_{k: k\in K_a\cap s} \frac{r_{uv}^{(k)}}{B_u^{(s)}}, \label{eq:thm1phi1} 
	\end{alignat}
where the last inequality follows since $\eta_u^{(s)} = \sum_{k: k\in K_a\cap s} \frac{r_{uv}^{(k)}}{B_u^{(s)}}$ 
and $a^x \leq 1 + ax \text{ for } 0 \leq x \leq 1, a \geq 1$. By rearranging the RHS of inequality \eqref{eq:thm1phi1}, 
we have:
\begin{equation*} 
	\sum_s \Delta \phi_u^{(s)} 
	\leq 2\lg(2p + 2)\cdot \sum_{k\in K_a} r_{uv}^{(k)} \cdot \left(\sum_{s: k\in s} 
		 \frac{\phi_u^{(s)}}{B_u^{(s)}} + \frac{|\{s: k\in s\}|}{p} \right)
	\leq 4\lg(2p + 2)\cdot  \sum_{k\in K_a} r_{uv}^{(k)},
\end{equation*} 
since $k$ is active and $p\geq |\{s: k\in s\}|$.
%
%
\end{proof}

A competitive ratio of $O(\lg p)$ for \gen in the small bids case now follows from 
Lemmas~\ref{lma:feasible}, \ref{lma:loss}, and \ref{lma:pot}, completing proof of the upper bound in Theorem \ref{thm:gen}.

%% file: ec-mm-gen-moderate-lb.tex

%

\section{\genaon lower bound}
\label{subsec:moderate-lb} 

In this section, we prove the lower bound for Theorem \ref{thm:genaon}. 
As a byproduct of this lower bound, we also obtain tight bounds for the online admission control problem, 
slightly improving the classical bounds of [Awerbuch et al. 1993].

Recall that we assume $\epsilon > \frac{1}{\lg(2p)}$.
In our instance, we have a single advertiser with a 
set of $p$ budget constraints (we index budget constraints
by $s$), where each constraint has a budget of 1. We will 
assume that $p^{\frac{\epsilon}{1-\epsilon}}$ 
is integral. (Note that $\epsilon > \frac{1}{\lg (2p)}$
implies that  $p^{\frac{\epsilon}{1-\epsilon}} > 2$.) Each impression has
bid value of $\epsilon$ on a single, unique dimension. 
(This allows us to use impressions and dimensions 
interchangeably in the rest of the construction.)
We are now left to specify the mapping of impressions 
to constraints. Let us denote $\ell = \frac{1-\epsilon}{\epsilon}$. 
First, we construct a hierarchical segmentation 
of the constraints $1, 2, \ldots, p$, where segment $j$ of 
level $i$ comprises constraints 
$(j-1) \cdot p^{1 - i/\ell} + 1, (j-1) \cdot p^{1 - i/\ell} + 2, 
\ldots, j\cdot p^{1 - i/\ell}$. Overall,
there are $\ell + 1$ levels $i = 0, 1, \ldots, \ell$, and 
level $i$ partitions the overall set of $p$ constraints into 
$p^{i/\ell}$ segments $j = 1, 2, \ldots, p^{i/\ell}$. Each 
such segment comprises $p^{1 - i/\ell}$ 
constraints. Note that the segments in level $i$ are a 
refinement of the segments in level $i-1$. 
For the purpose of visualization, 
the reader can imagine a complete $p^{1/\ell}$-ary tree on $p$
leaves, where each leaf represents a constraint and each internal node 
corresponds to a segment. 

The online arrival of impressions is divided into $\ell + 2$ 
rounds. The first round is special and is called the 
{\em initial round} (described below).
Every subsequent round, indexed $0, 1, \ldots, \ell$, 
comprises a set of {\em impression blocks}.
Each impression block of round $i$ corresponds to a unique segment 
in level $i$. Such an impression block comprises $1/\epsilon$
identical impressions, all of which appear in all constraints 
of the segment of level $i$ that the impression block 
corresponds to. Clearly, the total number 
of impression blocks in round $i$ is at most the total number 
of segments in level $i$, i.e., at most $p^{i/\ell}$. However,
not all segments receive an impression block corresponding
to it; only the {\em active} segments (we define this notion 
below) in level $i$ receive an impression block. Therefore, 
the arrival rule for round $i$ is simple: every active segment 
in level $i$ receives a unique impression block. The relative 
order of arrival of impressions in a round is arbitrary.  
As we mentioned above, the initial round is special --- in this
round, there is a single impression with bid value $\delta > 0$ 
that appears in all constraints. 

Now, we are only left to describe the rule for defining active
segments. After the initial round,
all segments in all levels are inactive, except 
the single segment in level 0 comprising all dimension. Therefore, 
in round 0, there is a single impression block corresponding 
to all the constraints. Next, we define the inductive process
for activating segments. Recall that the impression blocks
arriving in round $i \geq 0$ correspond to the active 
segments in level $i$. The algorithm assigns some subset 
of these impressions to the sole advertiser. For any segment
$j$ in level $i$ that received an impression block, 
if the algorithm assigned $t > 0$ impressions from the
block, then all impression blocks at level $i+t$ that 
are refinements of the current block are made active.
After round $i$, all segments in level $i$ are made inactive.
This completes the description of the instance.

First, we need to show that the construction is valid. In 
particular, we need to show that $i+t \leq \ell$ (i.e., 
there is a level where the refined segments can be activated) 
if the algorithm assigns $t$ impressions of a segment in round 
$i$. We prove a more general lemma.
\begin{lemma}
\label{lma:active-new}
	At any point of time, if a segment in level $i$ is active, then
	every constraint in that segment has a current utilization of 
	$\epsilon i + \delta$.
\end{lemma}
\begin{proof}
	The algorithm must assign the impression in the initial round to 
	stay competitive; therefore, 
	the lemma holds for $i=0$. Inductively, the 
	segments that are made active at level $i+t$ had utilization
	$\epsilon i$ before round $i$ (by the inductive hypothesis) and 
	have an additional utilization of $\epsilon t$ from round $i$.
\end{proof}
As a corollary of this lemma, we can infer that $i+t \leq \ell$
since utilization cannot exceed 1; hence, the construction is valid. 

Let $\ell_s$ be the last round where constraint $s$ was in an active 
segment, and let the corresponding segment be $\tau_s$. 
For every constraint $s$, the optimal solution assigns the entire 
impression block corresponding to $\tau_s$ in level $\ell_s$. 
Note that the segments $\tau_s$ partition the set of constraints, 
and hence, the optimal assignment is feasible. Clearly, the 
optimal solution earns a revenue of 1 on the impression block
corresponding to segment $\tau_s$. 

To compare the revenue of 
the algorithm, we redistribute the revenue earned by the algorithm 
on an impression by dividing it equally among all the constraints
that the impression appears in. Next, we sum the revenues on 
constraints in the same segment $\tau_s$. We will now compare this
revenue on $\tau_s$ with the unit revenue that the optimal solution
earns. To upper bound the (redistributed) revenue of the algorithm,
we first note that the algorithm does not earn any revenue on
the impression block $\tau_s$ itself in round $\ell_s$. If 
$\ell_s \leq \ell - 1$, this follows from the fact that 
refinements of $\tau_s$ were not marked active. If $\ell_s = \ell$, 
then by Lemma~\ref{lma:active-new}, the utilization at the beginning of 
the round is $1 - \epsilon + \delta$, which prevents the algorithm 
from earning any further revenue (note that revenue comes in units
of $\epsilon$).
\begin{lemma}
\label{lma:algo-ub}
	The total revenue earned by segment $\tau_s$ in the algorithm 
	is at most $\frac{2 \epsilon}{p^{1/\ell}}$.
\end{lemma}
\begin{proof}
	Let $r_0, r_1, \ldots, r_j$ denote the rounds in which segment
	$\tau_s$ earns revenue, and let $t_0, t_1, \ldots, t_j$ be the 
	number of impressions assigned in the respective rounds. In 
	other words, $r_0 = 0, r_1 = t_0 + r_0, r_2 = t_1 + r_1, \ldots, 
	\ell_s = r_j + t_j$. Then, the revenue earned by constraint $s$ is
	\begin{equation*}
		\sum_{q=0}^{j} \epsilon t_q\cdot \frac{1}{p^{1-r_q/\ell}} 
		\leq \epsilon \cdot \sum_{i=0}^{\ell_s - 1} \frac{1}{p^{1-i/\ell}} 
		\leq \frac{2\epsilon}{p^{1-(\ell_s - 1)/\ell}},	\end{equation*}
	where the last inequality follows $\text{since } p^{-1/\ell} \leq 1/2 \text{ for } \ell \leq \lg p \text{ which follows from } \epsilon > \frac{1}{\lg (2p)}$.
	Since $\tau_s$ comprises $p^{1 - \ell_s/\ell}$ constraints,
	the revenue earned by the algorithm on segment $\tau_s$
	is at most $\frac{2 \epsilon}{p^{1/\ell}}$.
\end{proof}
%
%
The lower bound now follows by comparing Lemma~\ref{lma:algo-ub} to the
optimal revenue, and setting $\delta \rightarrow 0$.

%% file: ec-mm-gen-small-lb.tex
\section{\gen and \genp lower bound (Theorem \ref{thm:gen} and Theorem \ref{thm:genp}) }
\label{subsec:gen-sb} 

In this section, we prove a $\Omega(\lg p)$ lower bound for \gen that will be constructed via a reduction from the online problem considered  by Awerbuch et al.\ in \cite{AwerbuchAP93}, 
which we call {\sc admission-control}. At the end of the section, we will describe how the problem definition of {\sc admission-control} can be modified so 
that the reduction implies the same lower bound for \genp. 
We formally define this problem as follows. \\ 

{\sc admisison-control:} At the outset, the online algorithm is given an edge-capacitated graph $G = (V, E)$, where $c_e$ denotes the capacity of edge $e$.
Demand requests then arrive in an online sequence $R = \langle r_1, \ldots, r_h \rangle$, where each request $r_i$ is specified by a path $P_i$ between two vertices $(s_i, t_i)$ and a
capacity demand $d_i$.  Upon the arrival of $r_i$, the algorithm must decide to either reject the request or route it
 along $P_i$.  The objective of the algorithm is to maximize the total demand of admitted requests subject 
to the constraint that for any edge $e$, the sum of capacity demands from admitted requests using $e$ does not exceed $c_e$. \\

We note that this is a less general problem than the one considered in \cite{AwerbuchAP93} 
(e.g.\ in the original problem, the algorithm can choose routing paths, requests have arrival times, etc.); however to extend their lower bound, 
the above problem definition will suffice. 

For completeness, we will give the proof of the ``small-demands" {\sc Admission-Control} lower bound from  \cite{AwerbuchAP93}
(i.e. in the given instance, the maximum demand-to-capacity ratio is arbitrarily small). 
This will be useful as our reduction will not be completely ``black-box", i.e., 
we need to have some knowledge of the proof's online sequence in order to perform the reduction. 

Let $L(n) $ be the line graph defined on $n+1$ vertices $\{v_1, \ldots, v_{n+1}\}$ (with $n$ edges). Then the following lemma holds. 

\begin{lemma} 
\label{lem:sbpathlb}
(\cite{AwerbuchAP93})
Let $A$ an online algorithm for {\sc Admission-control} under the small demands assumption.  Then there exists an instance $I$ for $L(n)$ 
such that the capacity earned by the optimal solution is $\Omega(\lg n)$ times the demand earned by $A$.  
\end{lemma} 

\begin{proof} 

Without loss of generality, assume $n$ is a power of two. 
The instance will consist of $\lg n +1$ phases indexed by $i = 0, \ldots, \lg n$. In phase $i$, we will issue $2^i$ groups of requests. 
The $j$th group in phase $i$ consists of $1/\delta$ requests each with capacity demand $\delta$ and identical routing paths $P_j$ (and 
so the total capacity for each group in every phase is 1). 

Fix a phase $i$ and group $j \in \{0, 1, \ldots, 2^i -1\}$.  Then path $P_j$ is defined to be the segment of $L(n)$ starting at vertex
$v_{\frac{jn}{2^{i}}}$ and ending at vertex $v_{\frac{(j+1)n}{2^i}}$. In other words, in a given phase we are splitting $L(n)$ into 
$2^i$ edge -disjoint subsegments each of the length $n/2^{i}$, where each subsegment defines a path for a group.  

Let $x_i$ be the total demand admitted by the algorithm from requests in phase $i$. Observe that in order to admit a unit of demand from the 
requests in phase $i$, the algorithm must use up a total capacity of $n/2^{i}$ (since each path $P_j$ in phase $i$ has $n/2^{i}$ edges). 
Since the total capacity of edges in the graph is $n$, we have that $\sum_{i = 0}^{\lg n} 2^{-i}nx_i  \leq n$. This implies that 
\begin{equation} 
\label{eq:pccapbound} 
\sum_{i = 0}^{\lg  n} 2^{-i}x_i \leq 1. 
\end{equation} 

Let $S_k = 2^{-k}\sum_{i = 0}^{k} x_i$ be the normalized total capacity obtained by the algorithm from phases 0 through $k$.
By equation \eqref{eq:pccapbound}, we have that 

\begin{equation*} 
\sum_{k = 0}^{\lg n} S_k = \sum_{k = 0}^{\lg n}\sum_{i = 0}^{k} 2^{-k}x_i \leq 2\sum_{k = 0}^{\lg n} 2^{-k}x_k \leq 2.
\end{equation*} 
Thus, there exists a $k'$ such that $S_{k'} \leq 2/\lg n$, 
implying the algorithm only earned at most $2^{k'+1}/\lg n$ total capacity after the completion of phase $k'$. The optimal solution 
at this point is to reject all requests before phase $k'$ and admit all requests in phase $k'$ to obtain capacity $2^{k'}$. Therefore, the 
adversary can stop the sequence after phase $k'$ to obtain the desired instance.
\end{proof}

Given Lemma \ref{lem:sbpathlb} and its proof,  let $I$ be the instance implied by the statement of the lemma, and let $I'$ be the entire of sequence 
of requests given by the construction (all $\lg n +1$ phases, regardless of the algorithm's behavior).  We construct our lower bound instance for \genaon (in the same bids case) as follows.

\begin{itemize} 
\item There will be one bidder $u$ for the instance. 
\item Each group in $I'$ will correspond to both a dimension and an impression. Specifically, 
for all impressions that correspond to the $v$th group in $I'$, we set $r_{uv}^{(v)} = \delta$  and $r_{uv}^{(k)} = 0$ for all $k \neq v$.
\item Each edge $e$  will correspond to a budget constraint $B_{u}^{(s_e)}$, where the capacity of the budget is $c_e =1$ and set $s_e$ is 
defined to be the set of dimensions whose corresponding request group use edge $e$ along their paths in $I'$.  
\end{itemize}

Observe that the single request in phase 1 of $I'$ traverses all of $L(n)$; therefore based on the construction, the first dimension belongs to $n$ different budget constraints, 
which implies $p = n$. It now follows from Lemma \ref{lem:sbpathlb} that any algorithm for \genaon is $\Omega(\lg p)$ competitive. \\ 

\noindent {\bf Extension to \genp:} To show a $\Omega(\lg p)$ lower bound for \gen (without the small bids assumption),
 we modify the definition of {\sc Admission-control} so that algorithm chooses to accept each 
request with some fraction $f_i \in [0,1]$ (i.e., the algorithm routes demand $f_i\cdot d_i$ for request $r_i$). We then change the lower bound instance in  Lemma \ref{lem:sbpathlb} so that 
each request has unit demand (instead of issuing $1/\delta$ requests with $\delta$ demand). The remainder of the reduction is equivalent. Using the same arguments as before, we obtain the desired $\Omega(\lg p)$ lower bound 
for the \genp setting.

%% file: ec-mm-gen-moderate-algo.tex


\section{\genp and \genaon upper bound (Theorem \ref{thm:genp} and Theorem \ref{thm:genaon})}

In this section, we give our upper bounds for \genp and \genaon.
We will first present the algorithm under the context \genaon. At the end of the section, we will outline how the same 
analysis extends to \genp. In both settings, the algorithm and analysis will be almost identical to our 
algorithm for \gen in Section \ref{sec:gen}.
\subsection{Algorithm Definition } 
\label{subsec:moderate-ub} 

Again let $\kappa_u^{(s)}$ denote the fraction of $B_u^{(s)}$ currently used by the algorithm.  
The algorithm (we call it \algo) uses an exponential potential function defined by:
\begin{equation*}
	\phi = \sum_u \sum_s \phi_u^{(s)} = \sum_u \sum_s \frac{B_u^{(s)}}{p} \left((p + 1)^{\frac{\kappa_u^{(s)}}{1-\epsilon}} - 1\right).
\end{equation*}
Note that $\phi = 0$ initially.

At any stage of \algo, a dimension $k$ is said to be 
	{\em active} for bidder $u$ if and only if $\sum_{s: k\in s} \frac{\phi_u^{(s)}}{B_u^{(s)}} \leq 1$, 
	else dimension $k$ is said to be {\em inactive} for bidder $u$. 
	(Note that this is a different definition of active dimensions than what is used in Section \ref{sec:lam-sb}). 
\algo only attempts to earn revenue on active dimensions, and hence, the total revenue if impression 
$v$ is allocated to bidder $u$ is given by:
\begin{equation*}
	r_{uv} = \sum_{k\in A_u} r_{uv}^{(k)}, \text{ where } A_u \text{ is the set of current active 
		dimensions for bidder } u.
\end{equation*}
The algorithm makes a greedy assignment with respect to $r_{uv}$, i.e., it assigns impression $v$ to 
$\arg\max_{u} r_{uv}$. Note that it is possible that $A_u = \emptyset$ for all bidders $u$,
and therefore the algorithm does not assign impression $v$ to any bidder, even though there are 
dimensions and bidders where it could have earned revenue. This completes the description of our algorithm. \\ 


\subsection{Algorithm Analysis}
 We first establish feasibility of the solution.
\begin{lemma}
\label{lma:feasible-mod-app}
	If \algo assigns an impression $v$ to a bidder $u$, then it can earn revenue on all the 
	active dimensions $A_u$ of $u$ without violating any constraint.
\end{lemma}
\begin{proof}
	We need to show that for all constraints $s$ of bidder $u$,
		$\kappa_u^{(s)} + \sum_{k\in A_u\cap s} \frac{r_{uv}^{(k)}}{B_u^{(s)}} \leq 1$.
	Suppose not. Then, for some constraint $s$,
	\begin{eqnarray*}
		\kappa_u^{(s)} + \epsilon
		\quad \geq \quad \kappa_u^{(s)} + \sum_{k\in A_u\cap s} \frac{r_{uv}^{(k)}}{B_u^{(s)}} 
			& > & 1 \quad (\text{by the definition of } \epsilon)\\
		\text{i.e., } (p+1)^{\frac{\kappa_u^{(s)} + \epsilon}{1-\epsilon}} & > & (p + 1)^{\frac{1}{1-\epsilon}} \\
		\text{i.e., } (p + 1)^{\frac{\kappa_u^{(s)}}{1-\epsilon}} & > & p + 1 
		\quad \left(\text{since } \frac{\kappa_u^{(s)}}{1-\epsilon} > 1 \text{ from the first line}\right)\\
		\text{i.e., } \frac{p \phi_u^{(s)}}{B_u^{(s)}} + 1 & > & p + 1 \\
		\text{i.e., } \frac{\phi_u^{(s)}}{B_u^{(s)}} & > & 1, 
	\end{eqnarray*}	
	which contradicts the fact that dimension $k$ is active for bidder $u$.
\end{proof}
This lemma implies that \algo is indeed able to earn revenue on all active dimensions of a bidder $u$ when 
it assigns an impression to $u$.

Next, we will bound the total revenue of an optimal solution that we denote by \opt against the total revenue
of \algo. Let $u_{\opt}(v)$ (resp., $u_{\algo}(v)$) be the bidder 
that \opt (resp., \algo) allocates impression $v$ to. For every dimension $k$ that \opt earns revenue from, 
one of the following holds:
\begin{enumerate}
	\item {\bf Case 1:} dimension $k$ is active for bidder $u_{\opt}(v)$ in \algo when impression $v$ arrives, and \algo 
		assigns $v$ to the same bidder, i.e., $u_{\algo}(v) = u_{\opt}(v)$.
	\item {\bf Case 2:} dimension $k$ is active for bidder $u_{\opt}(v)$ in \algo when impression $v$ arrives, but \algo 
		assigns $v$ to a different bidder, i.e., $u_{\algo}(v) \not= u_{\opt}(v)$.
	\item {\bf Case 3:} dimension $k$ is inactive for bidder $u_{\opt}(v)$ in \algo when impression $v$ arrives.
\end{enumerate}
We partition the dimensions that \opt earns revenue from into active and inactive dimensions (according
to their status in \algo for bidder $u_{\opt}(v)$). For active dimensions (cases 1 and 2 above), the next lemma gives a 
straightforward charging argument using the greediness of the choice made by \algo.

\begin{lemma}
\label{lma:active-mod-app}
	For any impression $v$, the total revenue earned by \opt on the active dimensions is at most the total 
	revenue earned by \algo overall.
\end{lemma}
\begin{proof}
	For case 1 above (\opt and \algo choose the same bidder), the two algorithms earn the same revenue on
	the active dimensions. For case 2 above, the fact that \algo makes a greedy choice implies that it earns
	at least as much revenue by assigning to a different bidder as it would have made by assigning to $u$,
	which includes the revenue on all the active dimensions.
\end{proof}	

The more involved case is that of inactive dimensions (case 3 above). In this case, we use a different global charging 
argument over all dimensions, based on the potential function. In particular, we show that for a bidder 
$u$, the total revenue of \opt from inactive dimensions (recall that this only includes revenue from 
impressions that arrived {\em after} the dimension became inactive in \algo) can be charged, up to a 
loss equal to the desired competitive ratio, to the revenue that \algo earned overall from bidder $u$.

\begin{lemma}
\label{lma:loss-mod-app}
	Fix a bidder $u$.  The total revenue that \opt earns in inactive dimensions for bidder $u$ is at most 
	the final potential of bidder $u$ in \algo.
\end{lemma}	
\begin{proof}
	For any dimension $k$, let $V_{u, k}$ denote the subset of impressions assigned to $u$ by \opt 
	that arrived after $k$ became an inactive dimension for bidder $u$ in \algo. We need to bound the 
	total revenue earned by \opt on dimension $k$ from impressions in $V_{u, k}$, summed over all $k$. 
	For any impression $v\in V_{u, k}$, we have:
	\begin{equation*}
		\sum_{s: k\in s} \frac{\phi_u^{(s)}}{B_u^{(s)}} > 1,
	\end{equation*}
	where $\phi_u^{(s)}$ is the final potential for constraint $s$ of bidder $u$.
	Thus,	the revenue that \opt earns from impressions $v\in V_{u, k}$ on dimension $k$,
	summed over all dimensions, can be bounded as follows:
	\begin{equation*}
		\sum_k \sum_{v\in V_{u, k}} r_{uv}^{(k)}
		< \sum_k \sum_{v\in V_{u, k}} r_{uv}^{(k)} \sum_{s: k\in s} \frac{\phi_u^{(s)}}{B_u^{(s)}} 
		= \sum_s \phi_u^{(s)} \sum_{k\in s} \sum_{v\in V_{u, k}} \frac{r_{uv}^{(k)}}{B_u^{(s)}} 
		\leq \sum_s \phi_u^{(s)},
	\end{equation*}	
where the last inequality follows by the feasibility of {\sc OPT}. 
\end{proof}

Next, we need to lower bound the final potential of a bidder in terms of the revenue that 
\algo earns from her. This is given in the next lemma.
\begin{lemma}
\label{lma:pot-mod-app}
	The increase in potential of a bidder $u$ during the course of $\algo$ is at most $4 \cdot \frac{p^{\frac{\epsilon}{1-\epsilon}}}{\epsilon}$ 
	times the revenue that \algo earns from $u$.
\end{lemma}
\begin{proof}
	Suppose \algo assigns impression $v$ to bidder $u$. Let $K_a$ denote the set of active dimensions
	for bidder $u$ when this assignment is made. Let $\eta_u^{(s)} = \sum_{k: k\in K_a\cap s} \frac{r_{uv}^{(k)}}{B_u^{(s)}}$. 
	The increase in potential for bidder $u$ is given as follows: 
	
\begin{alignat}{2} 
       \sum_s \Delta \phi_u^{(s)} 
       & = \sum_s \frac{B_u^{(s)}}{p} \left((p+1)^{\frac{\kappa_u^{(s)} + \eta_u^{(s)}}{1-\epsilon}} - (p+1)^{\frac{\kappa_u^{(s)}}{1-\epsilon}}\right) \notag\\
	 & = \sum_s \frac{B_u^{(s)}}{p} (p+1)^{\frac{\kappa_u^{(s)}}{1-\epsilon}} \left((p+1)^{\frac{\eta_u^{(s)}}{1-\epsilon}} - 1\right)  \notag \\
	 & = \sum_s \left(\phi_u^{(s)} + \frac{B_u^{(s)}}{p}\right) \cdot \left(\left((p+1)^{\frac{\epsilon}{1-\epsilon}}\right)^{\frac{\eta_u^{(s)}}{\epsilon}} - 1\right) \notag \\
	& \leq \sum_s \left(\phi_u^{(s)} + \frac{B_u^{(s)}}{p}\right) \cdot \frac{(p+1)^{\frac{\epsilon}{1-\epsilon}}}{\epsilon} 
	   \cdot \sum_{k: k\in K_a\cap s} \frac{r_{uv}^{(k)}}{B_u^{(s)}}, \label{eq:thm1phi1-app} 
	\end{alignat}
where the inequality follows since $\eta_u^{(s)} = \sum_{k: k\in K_a\cap s} \frac{r_{uv}^{(k)}}{B_u^{(s)}} \leq \epsilon$ 
and $a^x \leq 1 + ax \text{ for } 0 \leq x \leq 1, a \geq 1$. By rearranging the RHS of inequality \eqref{eq:thm1phi1-app} and using the fact that $|\{s: k\in s\}| \leq p$, we obtain:

\begin{alignat*}{2} 
	\sum_s \Delta \phi_u^{(s)} 
	& \leq \frac{(p+1)^{\frac{\epsilon}{1-\epsilon}}}{\epsilon} \sum_{k\in K_a} r_{uv}^{(k)} \cdot \left(\sum_{s: k\in s} 
		 \frac{\phi_u^{(s)}}{B_u^{(s)}} + \frac{|\{s: k\in s\}|}{p} \right)\\ 
	& \leq  2 \cdot \frac{(p+1)^{\frac{\epsilon}{1-\epsilon}}}{\epsilon} \sum_{k\in K_a} r_{uv}^{(k)} 
		\hspace{12mm}  \left(\text{since } k \text{ is active}, \text{ and } p\geq |\{s: k\in s\}|\right) \\
	& \leq 4 \cdot \frac{p^{\frac{\epsilon}{1-\epsilon}}}{\epsilon} \sum_{k\in K_a} r_{uv}^{(k)}
		\hspace{20mm} \left(\text{since } (p+1)^{\frac{\epsilon}{1-\epsilon}} \leq 2\cdot p^{\frac{\epsilon}{1-\epsilon}} \text{ for large enough } p\right),
\end{alignat*} 
as desired. 
\end{proof}
A competitive ratio of $O\left(\frac{p^{\frac{\epsilon}{1-\epsilon}}}{\epsilon}\right)$ for \genaon now follows from 
Lemmas~\ref{lma:active-mod-app}, \ref{lma:loss-mod-app}, and \ref{lma:pot-mod-app}. \\ 

\subsection{Extension to \genp} 

We begin by noting that for \genp, we will assume that the maximum bid-to-budget ratio 
$\max_{u, v, s} \frac{\sum_{k\in K_s} r_{uv}^{(k)}}{B_u^{(s)}} < 1$. Obviously, this assumption 
is wlog for \genaon since any $u,v$ pair that results in $\epsilon > 1$ cannot be assigned (and it is easy to show an
arbitrarily large lower bound when $\epsilon = 1$). For \genp, however, allowing instances where $\epsilon > 1$ 
still admits a nontrivial problem definition since the algorithm can choose to earn partial revenues. However, since such a scenario
would clearly never arise in practice (i.e, an impression generating more revenue than a budget) and only complicates the analysis, 
we proceed with this added assumption.

To adapt our \genaon algorithm and analysis for \genp, the algorithm will now choose to earn $\frac{1}{\lg(2p + 2)}$ 
fraction of all revenues and set the parameter $\epsilon = \frac{1}{\lg(2p + 2)}$ in algorithm and proof
 (so essentially the algorithm treats the instance as if its a \gen small-bids instance). 
Otherwise, the algorithm behaves identically as before.  Since we are assuming the maximum bid-to-budget ratio is at most 1,
 this scaling procedure ensures that the revenue generated 
by an impression never increases the utilization of a constraint by more than a $\frac{1}{\lg(2p + 2)}$ factor. 

To show a $O(\lg p)$ competitive ratio,
it suffices to show that Lemmas~\ref{lma:feasible-mod-app} through \ref{lma:pot-mod-app} still hold in this setting. It is not too hard to verify that Lemmas \ref{lma:feasible-mod-app}, 
\ref{lma:loss-mod-app}, and \ref{lma:pot-mod-app}  follow by the same arguments, noting that in these proofs, 
$r_{uv}^{(k)}$ still denotes the revenue earned by the algorithm after its been reduced by a $\frac{1}{\lg(2p + 2)}$ factor (except in Lemma \ref{lma:loss-mod-app}, $r_{uv}^{(k)}$ denotes 
the amount of revenue the optimal solution chooses to earn).  Lemma \ref{lma:active-mod-app} uses the same argument, except now the greedy property 
implies that the algorithm earns at least a $\frac{1}{\lg(2p + 2)}$ factor of that earned by the optimal solution on a active dimension (instead of strictly more); however, losing 
this factor in this case is fine since the we are ultimately aiming for a $O(\lg p)$ competitive ratio. Hence, our algorithm extends to the \genp setting.